\documentclass[3p,times]{elsarticle}

\usepackage{ecrc}

\usepackage{graphicx}
\usepackage{dcolumn}
\usepackage{bm}
\usepackage{epstopdf}
\usepackage{mathrsfs}
\usepackage{amssymb}
\usepackage{amsmath}

\def\q{{\boldsymbol q}}

\def\k{{\boldsymbol k}}

\def\x{{\boldsymbol x}}

\def\bkappa{{\boldsymbol \kappa}}
\def\bbkappa{\bar{\boldsymbol \kappa}}

\def\sM{\text{med}}

\def\pk{p\cdot k}
\def\hk{\bar{p} \cdot k}

\def\bbl{\bar{\boldsymbol{L}}}
\def\bc{\boldsymbol{C}}

\newcommand{\beq}{\begin{equation}}
\newcommand{\eeq}{\end{equation}}
\newcommand{\bes}{\begin{subequations}}
\newcommand{\ees}{\end{subequations}}
\newcommand{\bal}{\begin{align}}
\newcommand{\eal}{\end{align}}
\newcommand{\be}{\begin{eqnarray}}
\newcommand{\ee}{\end{eqnarray}}

\volume{00}

\firstpage{1}
\journalname{Nuclear Physics A}

\runauth{}

\jid{nupha}

\usepackage{amssymb}
\usepackage[figuresright]{rotating}
\begin{document}

\begin{frontmatter}

\dochead{}

\title{Coherence and broadening effects in medium induced gluon radiation}

\author[usc]{N\'estor Armesto}
\author[usc]{Hao Ma}
\author[usc]{Mauricio Martinez}\ead{mauricio.martinez@usc.es}
\author[saclay]{Yacine Mehtar-Tani}
\author[usc,cern]{Carlos A. Salgado}

\address[usc]{Departamento de F\'isica de Part\'iculas and IGFAE,
Universidade de Santiago de Compostela,
E-15782 Santiago de Compostela, 
Galicia-Spain.}
\address[saclay]{Institut de Physique Th\'eorique, CEA Saclay, F-91191 Gif-sur-Yvette, France.}
\address[cern]{Physics Department, Theory Unit, CERN, CH-1211 Gen\'eve 23, Switzerland.}

\begin{abstract}
Interferences between different emitters in the multi-parton shower is the building block of QCD jet physics in vacuum. The presence of a hot medium made of quarks and gluons is expected to alter this interference pattern. To study such effects, we derive the gluon emission spectrum off an "asymptotic quark" traversing a hot and dense QCD medium at first order in the medium density. The resulting induced gluon distribution gets modified when the new interference terms between the initial and final quark are included. We comment on the possible phenomenological consequences of this new contribution for jet observables in heavy-ion collisions.
\end{abstract}

\begin{keyword}
Perturbative QCD \sep jet physics \sep jet quenching 
\end{keyword}

\end{frontmatter}

\section{ Introduction}
\label{sec:intro}
Jets in hadronic collisions are an important tool to test experimentally many of the predictions based on pQCD. An striking property which determines the emission pattern of the observed gluon radiation in jets is the interference among different emitters in a parton cascade.  Color coherence effects have been verified experimentally in $e^+ e^-$  and $p \bar{p}$ collisions \cite{expcoherence}. The most remarkable consequence of color coherence phenomena in QCD is the angular ordering of subsequent soft gluon emissions which leads to the suppression of radiation at large angles. 

The suppression of soft radiation can be easily understood by considering gluon emission off in DIS with a highly virtual photon (t-channel). In this case, the single inclusive gluon spectrum is given by 
\beq
\label{vacspec}
\omega\frac{dN^\text{vac}}{d^3\vec{k}}=\frac{\alpha_s  C_F}{(2\pi)^2}\,\frac{p\cdot\bar{p}}{(\pk)(\hk)}\,,
\eeq
where $C_F$ is the color charge of the quark, $k=(k^+,k^-, {\bf k})$ is the 4th-momentum of the emitted gluon in light cone coordinates and $p(\bar{p})$ is the momentum of the in(out)coming parton\footnote{Hereafter an overlining $\bar{\ \ }$ is used to denote the momentum and related quantities of the outgoing quark.}. The contribution to the gluon spectrum (\ref{vacspec})  from either the incoming or the outcoming parton  can be found by separating the collinear divergences along the emitter and including the interferences. As a result, a probabilistic interpretation for a coherent gluon emission is found. The probability of emit a gluon from any of the emitters can be found by integrating along the azimuthal angle of one of the emitters, say the incoming parton  
\beq
\label{vacemiss}
\langle dN_{in}\rangle_{\phi}=\frac{\alpha_s C_F}{\pi} \frac{d\omega}{\omega} \frac{d(\cos\theta)}{1-\cos\theta}\Theta (\cos\theta-\cos\theta_{qq})\,,
\eeq
where $\theta$ is the angle of emitted gluon and $\theta_{qq}$ is the angle between the incoming and the outcoming parton due to hard scattering with the virtual photon. 
From last expresion we observe the presence of soft and collinear divergences which must be re-summed \cite{BCM}. In addition, Eq.(\ref{vacemiss}) indicates us that gluon radiation is going to be confined inside the cone defined by the opening angle $\theta_{qq}$ along the emitter. Hence, there is no large angle emission. 

Most of the efforts to understand medium induced gluon radiation have been limited to one inclusive gluon emission\cite{bdmps} but not much is known about color coherence effects in the presence of a QCD medium. Recently, this problem started to be addressed by some authors by considering the radiation pattern of a $q\bar{q}$ antenna immersed in a QCD medium  \cite{antenna}. In these proceedings we review our recent work \cite{tchannel} related with an extension of color coherence studies inside a QCD medium  to a space-like (t-channel) scattering process.  

\section{Interference between initial and final state radiation in a QCD medium}
\label{sec:int}

To study the initial state radiation in the presence of a QCD medium, we consider the medium induced gluon radiation of a parton created in the remote past which suffers a hard scattering and subsequently it passes through a QCD medium of finite size. 

At first order in the medium field $A^-_{\sM}(x^+,\q)$ one finds that the total scattering amplitude for gluon radiation off the incoming and the outcoming quark is \cite{tchannel}
\be
\label{eq:totmedampl}
{\cal M}^a_{\lambda}&=&2ig^2  \int \frac{d^2 \q}{(2\pi)^2}\,\int^{L^+}_{x_0^+} d x^+\, [T\cdot A^-_{\sM}(x^+,\q)]^{ab}\nonumber\\
&\times&\biggl\{
Q_{in}^b\,\frac{\bkappa-\q}{\bigl(\bkappa-\q\bigr)^2} 
-Q_{out}^b\, \left[\frac{\bar{\bkappa}-\q}{(\bbkappa-\q)^2}\biggl[1- \exp\biggl(i \frac{(\bbkappa-\q)^2}{2 k^+}x^+\biggr)\biggr] +\frac{\bbkappa}{\bbkappa^2}\exp\biggl(i \frac{(\bbkappa-\q)^2}{2 k^+}x^+
\biggr) \right] \biggr\}\,,
\ee
where $Q_{in(out)}^b$ is the color charge of the in(out)coming parton, $\bkappa={\bf k}-x{\bf p}$ is the transverse momentum of the gluon relative to one of the incoming parton (similar definition for $\bbkappa$) and $L^+$ is the length of the medium.  The first term of the last expression corresponds to the reshuffling of the gluon emission off the incoming parton after its encountering with the color charges of the medium. The second and third terms are associated with gluon emission off the outcoming parton. The second one is interpreted as the interaction of the emitted gluon with the medium while the third one is the interaction of the quark before bremstrahlung emission of the gluon. 

After taking the square of the scattering amplitude (\ref{eq:totmedampl}), the medium induced gluon spectrum is given by \cite{tchannel}
\be
\label{eq:totmedspec}
\omega\frac{dN^\text{med}}{d^3\vec{k}} &=&\frac{4\,\alpha_s C_F \,\hat q}{\pi} \int \frac{d^2 \q}{(2\pi)^2}{\cal V}^2(\q)~\int_0^{L^+} dx^+\,\Biggl[\frac{1}{(\bkappa-\q)^2}-\frac{1}{\bkappa^2}\nonumber \\
&+&\,2\,\frac{\bbkappa\cdot\q}{\bbkappa^2(\bbkappa-\q)^2}\biggl(1-\cos\biggr[\frac{(\bbkappa-\q)^2}{2 k^+}x^+\biggr]\biggr)\,\nonumber\\
&+&\,2\,\Biggl\{\frac{\bkappa\cdot\bbkappa}{\bkappa^2\bbkappa^2}-\frac{\bbkappa\cdot(\bkappa-\q)}{\bbkappa^2(\bkappa-\q)^2}\Biggr\}
\,+\, 2 \Biggl\{\frac{\bbkappa\cdot(\bkappa-\q)}{\bbkappa^2(\bkappa-\q)^2} -\frac{(\bbkappa-\q)\cdot(\bkappa-\q)}{(\bbkappa-\q)^2(\bkappa-\q)^2}\,
\,\Biggr\}\biggl(1-\cos\biggr[\frac{(\bbkappa-\q)^2}{2 k^+}x^+\biggr]\biggr)
\Biggr]\,,
\ee
The gluon spectrum is composed by three parts: (i) the gluon emission off the incoming parton (first line) which is the bremmstrahlung of an accelerated color charge which undergoes rescattering, (ii) gluon emission off outcoming quark (second line) is identified with the medium induced radiation of the N=1 opacity expansion (GLV spectrum) \cite{bdmps} and (iii) the interferences between the incoming and the outcoming parton (third line). In addition, the spectrum contains soft and collinear divergences. 

To shed light on the relevant scales of the problem which determine the physics, we study in the next subsections the coherent and incoherent limit as well as the soft sector of the gluon spectrum (\ref{eq:totmedspec}).

\subsection{Incoherent limit}
\label{subsec:incolimit}
For short formation times $\tau_f << L^+$, the phases cancel and the gluon spectrum (\ref{eq:totmedspec}) simplifies to  \cite{tchannel}
\beq
\label{eq:incohlimit}
\omega\frac{dN^\text{med}}{d^3\vec{k}}\biggl.\biggr|_{\tau_f\ll L^+} =\frac{4\,\alpha_s C_F \,\hat q}{\pi} \int \frac{d^2 \q}{(2\pi)^2}{\cal V}^2(\q)~\int_0^{L^+} dx^+ \biggl\{\bbl^2+\bc^2(\bkappa-\q)-\bc^2(\bkappa) \biggr\}\,,
\eeq
where we use the definition of the the Lipatov vertex $\bbl$ in the light cone gauge and the transverse emission current $\bc (\bkappa)$ respectively\footnote{An identical definition follows for $\bc (\bkappa-\q)$ by changing $\bkappa\to\bkappa-\q$.}
\beq
\bbl=\frac{\bar{\bkappa}-\q}{(\bbkappa-\q)^2}-\frac{\bar{\bkappa}}{\bbkappa^2} \,,\hspace{2cm}
\bc (\bkappa)= \frac{\bkappa}{\bkappa^2}-\frac{\bbkappa}{\bbkappa^2}.
\eeq
This result shows us the two main mechanisms of gluon radiation: The first term of Eq. (\ref{eq:incohlimit}) corresponds to the genuine medium-induced radiation of an asymptotic parton that suffers a scattering with the medium. The two last terms corresponds to bremsstrahlung associated to the hard scattering  followed by the radiated gluon suffering a classical  sequential process of rescattering with the medium. The latest two terms include interferences among the incoming and the outgoing quarks alike. This result is a generalization of the probabilistic interpretation of the incoherent limit of the GLV spectrum \cite{bdmps}. 

\subsection{Coherent limit}
\label{subsec:incolimit}
When one considers large formation times $\tau_f\gg L^+$, the gluon spectrum is given by \cite{tchannel}
\beq
\omega\frac{dN^\text{med}}{d^3\vec{k}}\biggl.\biggr|_{\tau_f \gg L^+}= \frac{4\,\alpha_s C_F \,\hat q}{\pi} \int \frac{d^2 \q}{(2\pi)^2}{\cal V}^2(\q)~\int_0^{L^+} dx^+\biggl\{
\frac{1}{(\bkappa-\q)^2}-\frac{1}{\bkappa^2} 
+ 2\frac{\bbkappa\cdot\bkappa}{\bbkappa^2\bkappa^2}
-2 \frac{\bbkappa\cdot(\bkappa-\q)}{\bbkappa^2(\bkappa-\q)^2} 
\biggr\}\,.
\label{eq:coherent}
\eeq
The first two terms in the last expression correspond to the reshuffling of the gluon emission off the incoming quark. The next two terms are interferences between the initial and final state. In addition, notice that even though the independent GLV gluon spectrum gets suppressed due to the LPM effect, some of the interferences among the initial and final parton remain. Notice that when Eq. (\ref{eq:coherent}) is integrated in $k_\perp$ and neglecting finite kinematics of the gluon, one recovers the result shown in Eq. (3.4) of \cite{Arleo}.

\subsection{Soft limit}
\label{subsec:incolimit}

In the soft limit $\omega\to 0$ the medium-induced gluon spectrum gets simplified to  \cite{tchannel}
\beq
\label{eq:softlimspec}
\omega\frac{dN^\text{med}}{d^3\vec{k}}\biggl.\biggr|_{\omega\to0} 
=\frac{\alpha_s C_F }{(2\pi)^2}\,\Delta_{med}\, \Biggl(2 \frac{\bbkappa\cdot\bkappa}{\bbkappa^2\bkappa^2}  -  \frac{1}{\bkappa^2}  \Biggr)\,,
\eeq
where $\Delta_{med}=\hat{q}L^+/m_D^2\approx L/\lambda$ that is the effective number of scattering centers. If now we consider within this limit the full gluon spectrum $dN^{tot}|_{\omegaº\to 0}=(dN^{vac}+dN^{med})|_{\omega\to 0}$, this can be written as \cite{tchannel}

\beq
\label{eq:fullspec}
\omega\frac{dN^\text{tot}}{d^3\vec{k}}\biggl.\biggr|_{\omega\to0} 
= \frac{\alpha_s C_F }{(2\pi)^2}\,\bigl({\cal P}_{in}^{tot}+{\cal P}_{out}^{tot}\bigr)\bigr|_{\omega\to0}\,,
\eeq
where 
\bes
\be
\label{eq:totcontra}
{\cal P}_{in}^{tot}&=& \bigl(1-\Delta_{med} \bigr) \Biggl( \frac{1}{\bkappa^2} - \frac{\bbkappa\cdot\bkappa}{\bbkappa^2\bkappa^2} \Biggr)  \,,\\
{\cal P}_{out}^{tot}&=& \Biggl(\frac{1}{\bbkappa^2}\,-\,\bigl(1-\Delta_{med}\bigr)\frac{\bbkappa\cdot\bkappa}{\bbkappa^2\bkappa^2}\Biggr) \,. \label{eq:totcontrb}
\ee
\ees

In the absence of a medium $\Delta_{med}\to 0$ and Eq. (\ref{eq:fullspec}) reduces to the vacuum radiation pattern (\ref{vacspec}). In the opposite case of an opaque medium $\Delta_{med}\to 1$, there is a significant reduction of soft gluon radiation from the incoming parton\footnote{Due to unitarity constraints $\Delta_{med}\le 1$. }.

The fact that the gluon radiation can be factorized in two components allows us to have a probabilistic interpretation. The first term ${\cal P}_{in}^{tot}$ is the coherent gluon emission off the incoming quark reduced by the probability $\Delta_{med}$ that the emitted gluon interacts with the QCD medium.  In addition, ${\cal P}_{in}^{tot}$ will keep the same angular ordering constraint as in the vacuum case. The second term ${\cal P}_{out}^{tot}$ accounts for the partial decoherence of the emitted gluon due to the scatterings with the QCD medium which share some similarities observed in the $q\bar{q}$ antenna case \cite{antenna}.  Such decoherence is measured by the probability of an interaction with the medium $\Delta_{med}$. As a consequence,  the medium opens the phase space for large angle emissions, a property called antiangular ordering \cite{antenna}. The soft limit (\ref{eq:fullspec}) remains valid for a range of finite gluon energies as far as $\omega\theta_{qq},\k\ll m_D$.

\section{Conclusion}
\label{sec:concl}

In this work we study the radiation pattern of the initial state radiation in the presence of a QCD medium by including interferences between the incoming and the outcoming parton.
We calculate the medium induced gluon spectrum of an asymptotic parton created in the remote past which suffers a hard collision and subsequently crosses a QCD medium. The gluon spectrum is composed of the independent gluon emissions associated to the incoming and outgoing parton as well as interference terms between both emitters. We study three asymptotic limits of the medium induced gluon spectrum: the incoherent, coherent and soft sector. In the incoherent limit $\tau_f\ll L^+$, a probabilistic interpretation for the gluon spectrum is found which resembles a similar structure observed for the GLV spectrum but now it is generalized by accounting interferences among the two emitters. For the coherent limit $\tau_f\gg L^+$ the medium induced gluon spectrum is reduced to the classical broadening contribution associated exclusively to the initial state and some of the interferences remain.  In the soft limit, the probabilistic interpretation for the radiation pattern of the full gluon spectrum has two main important features: (i)  gluon emissions from the initial state remain coherent but they are reduced by the probability to interact with the medium measured by  $\Delta_{med}\sim \hat{q}/m_D^2$  and (ii) the final state emissions decoheres due to interactions with the medium so antiangular ordering appears due to interferences between both emitters. 

 Our studies show how interferences indeed affect the angular distribution of the gluon radiation under the presence of a QCD medium. The results found in the soft limit hold for any $t$-channel exchange of a color singlet object independent of the  color representations as far as $\omega\theta_{qq},\k\ll m_D$. Our work is a first principles approach for the proper inclusion of coherence effects on observables sensitive to initial state radiation in high energy nuclear collisions. 

\noindent{{\bf Acknowledgements\ \ } This work is supported by European Research Council grant HotLHC ERC-2001-StG-279579; by Ministerio de Ciencia e Innovaci\'on of Spain under projects FPA2008-01177, FPA2009-06867-E and FPA2011-22776; by Xunta de Galicia (Conseller\'{\i}a de Educaci\'on and Conseller\'\i a de Innovaci\'on e Industria -- Programa Incite); by the Spanish Consolider-Ingenio 2010 Programme CPAN and by FEDER. C.A.S. is a Ram\'on y Cajal researcher. The work of Y.M.T. is supported by the European Research Council under the Advanced Investigator Grant ERC-AD-267258.}

\end{document}